\documentclass[sn-aps]{sn-jnl}

\usepackage[english]{babel}
\usepackage{graphicx} 
\usepackage{geometry}
\usepackage{multirow}%
\usepackage{amsmath,amssymb,amsfonts}%
\usepackage{amsthm}%
\usepackage{hyperref}
\usepackage{breakurl}
\usepackage{setspace}
\usepackage{enumitem}  
\usepackage{calc}  

\begin{document}

\title{Latin American Strategy Forum for Research Infrastructure (III LASF4RI Contribution) \vspace*{2mm} \\
 \textbf{The glue that binds us all -- Latin America and the Electron-Ion Collider} }

\author[1]{\fnm{A.~C.} \sur{Aguilar}}
\author[2]{\fnm{A.} \sur{Bashir}}
\author[3]{\fnm{J.~J.} \sur{Cobos-Mart\'inez}}
\author*[4]{\fnm{A.} \sur{Courtoy}}\email{aurore@fisica.unam.mx}
\author*[5]{\fnm{B.} \sur{El-Bennich}}\email{bennich@unifesp.br}
\author[6]{\fnm{D.} \sur{de Florian}}
\author[7]{\fnm{T.} \sur{Frederico}}
\author[8]{\fnm{V.~P.} \sur{Gon\c{c}alves}}
\author[9]{\fnm{M.} Hentschinski}
\author[10]{\fnm{R.~J.} \sur{Hern\'andez-Pinto}}
\author[11]{\fnm{G.}  \sur{Krein}}
\author[12]{\fnm{M.~V.~T.}  \sur{Machado}}
\author[13]{\fnm{J.~P.~B.~C.}  \sur{de~Melo}}
\author[7]{\fnm{W.}  \sur{de Paula}}
\author[14]{\fnm{R.}  \sur{Sassot}}
\author[15]{\fnm{F.~E.}  \sur{Serna}}
\author[16]{\fnm{Supporting authors outside Latin America: L.}  \sur{Albino}}
\author[17]{\fnm{I.}  \sur{Borsa}}
\author[18]{\fnm{L.} \sur{Cieri}}
\author[19]{\fnm{I.~M.} \sur{Higuera-Angulo}}
\author[20]{\fnm{J.}  \sur{Mazzitelli}}
\author[18]{\fnm{\'A.}  \sur{Miramontes}}
\author[21]{\fnm{K.}  \sur{Raya}}
\author[22]{\fnm{F.}  \sur{Salazar}}
\author[23]{\fnm{G.}  \sur{Sborlini}}
\author[24]{\fnm{P.}  \sur{Zurita}}

\affil[1]{\orgdiv{Instituto de F\'isica Gleb Wataghin}, \orgname{Universidade Estadual de Campinas}, \orgaddress{ \city{Campinas}, \postcode{13083-859}, \state{S\~ao Paulo}, \country{Brasil} }}
\affil[2]{\orgdiv{Instituto de F\'isica y Matem\'aticas}, \orgname{Universidad Michoacana de San Nicol\'as de Hidalgo}, \orgaddress{  \city{Morelia},  \state{Michoac\'an}, \postcode{58040}, \country{M\'exico}}}
\affil[3]{\orgdiv{Departamento de F\'isica}, \orgname{Universidad de Sonora}, Boulevard Luis Encinas J.~y Rosales, Hermosillo, Sonora 83000, M\'exico}
\affil[4]{\orgdiv{Instituto de F\'isica}, \orgname{Universidad Nacional Aut\'onoma de M\'exico}, Apartado Postal 20-364, 01000 Ciudad de M\'exico, M\'exico} 
\affil[5]{\orgdiv{Departamento de F\'isica}, \orgname{Universidade Federal de S\~ao Paulo}, Rua S\~ao Nicolau 210, 09913-030 Diadema, S\~ao Paulo, Brasil}
\affil[6]{\orgdiv{International Center for Advanced Studies (ICAS) and IFICI}, \orgname{Universidad Nacional de San Mart\'in},  25 de Mayo y Francia, Buenos Aires, Argentina}
\affil[7]{\orgdiv{Instituto Tecnol\'ogico de Aeron\'autica}, \orgname{Departamento de Ci\^encia e Tecnologia Aeroespacial}, 12228-900 S\~ao Jos\'e dos Campos, Brasil}
\affil[8]{\orgdiv{Instituto de F\'isica e Matem\'atica}, \orgname{Universidade Federal de Pelotas}, Caixa Postal 354, 96010-900 Pelotas, Rio Grande do Sul, Brasil}
\affil[9]{\orgdiv{Departamento de Actuaria, F\'isica y Matem\'aticas}, \orgname{Universidad de las Am\'ericas Puebla}, Ex-Hacienda Santa Catarina Martir S/N, San Andr\'es Cholula, 72820 Puebla, M\'exico}
\affil[10]{\orgdiv{Facultad de Ciencias F\'isico-Matem\'aticas}, \orgname{Universidad Aut\'onoma de Sinaloa}, Ciudad Universitaria, Culiac\'an, Sinaloa 80000, M\'exico}
\affil[11]{\orgdiv{Instituto de F\'isica Te\'orica}, \orgname{Universidade Estadual Paulista}, 01140-070 S\~ao Paulo, S\~ao Paulo, Brazil}
\affil[12]{\orgdiv{Instituto de F\'isica}, \orgname{Universidade Federal do Rio Grande do Sul}, Caixa Postal 15051,91501-970 Porto Alegre, Rio Grande do Sul, Brasil}
\affil[13]{\orgdiv{Laborat\'orio de F\'isica Te\'orica e Computacional}, \orgname{Universidade Cidade de S\~ao Paulo}, 01506-000 S\~ao Paulo, S\~ao Paulo, Brasil}
\affil[14]{\orgdiv{Departamento de F\'isica and IFIBA, Facultad de Ciencias Exactas y Naturales}, \orgname{Universidad de Buenos Aires}, Pabell\'on 1 (1428), Buenos Aires, Argentina}
\affil[15]{\orgdiv{Departamento de F\'isica}, \orgname{Universidad de Sucre}, Barrio Puerta Roja, Sincelejo 700001, Colombia} 
\affil[16]{\orgdiv{Departamento de Sistemas F\'isicos, Qu\'imicos y Naturales}, \orgname{Universidad Pablo de Olavide}, E-41013 Sevilla, Spain}
\affil[17]{\orgdiv{Institute for Theoretical Physics}, \orgname{University of T\"ubingen}, Auf der Morgenstelle 14, 72076 T\"ubingen, Germany}
\affil[18]{\orgdiv{Department of Theoretical Physics \& Instituto de F\'isica Corpuscular}, \orgname{Universitat de Val\`encia \& Consejo Superior de Investigaciones Cient\'ificas}, E-46980 Paterna, Val\`encia, Spain}
\affil[19]{\orgdiv{Thomas Jefferson National Accelerator Facility, 12000 Jefferson Ave., Newport News, VA 23606, United States}}
\affil[20]{\orgname{Paul Scherrer Institut}, CH-5232 Villigen PSI, Switzerland}
\affil[21]{\orgdiv{Department of Integrated Sciences and Center for Advanced Studies in Physics}, \orgname{Mathematics and Computation}, University of Huelva, E-21071 Huelva, Spain}
\affil[22]{\orgdiv{Institute for Nuclear Theory}, \orgname{University of Washington}, Seattle WA 98195-1550, USA}
\affil[23]{\orgdiv{Departamento de F\'isica Fundamental \& IUFFyM}, \orgname{Universidad de Salamanca}, 37008 Salamanca, Spain}
\affil[24]{\orgdiv{Departamento de F\'isica Te\'orica \& IPARCOS}, \orgname{Universidad Complutense de Madrid}, E-28040 Madrid, Spain\vspace*{0.7cm}}


\abstract{ The Electron-Ion Collider~(EIC), a next generation electron-hadron and electron-nuclei scattering facility, will be built at Brookhaven National Laboratory. The wealth of new data will 
shape research in hadron physics, from nonperturbative QCD techniques to perturbative QCD improvements and global QCD analyses, for the decades to come. With the present proposal, 
Latin America based physicists, whose expertise lies on the theory and phenomenology side, make the case for the past and future efforts of a growing community, working hand-in-hand 
towards developing theoretical tools and predictions to analyze, interpret and optimize the results that will be obtained at the EIC, unveiling the role of {\it the glue that binds us all}. 
This effort is along the lines of various initiatives taken in the U.S., and supported by colleagues worldwide, such as the ones by the EIC User Group which were highlighted during the 
Snowmass Process and the Particle Physics Project Prioritization Panel~(P5).  }


\maketitle


\section{Hadron Physics in the wake of the EIC}

It is nowadays firmly established that Quantum Chromodynamics (QCD) is the fundamental theory of strong interactions~\cite{Gross:2022hyw,Achenbach:2023pba,Fritzsch:2015jfa} that 
describes the interaction of colored degrees of freedom: quarks and gluons. The theory has been tested over several decades and continuous progress has been made in verifying its accuracy 
against experimental data. At the core of this theory, however, lies the phenomenon of color confinement which is still not understood. Confinement implies that quarks and gluons cannot be 
observed freely in Nature, as they are always spatially bound within hadrons. We know that confinement exists, and our own existence is its proof, but elucidating its mechanism is one 
of the most important problems in modern physics and science~\cite{Accardi:2012qut}. Understanding this fundamental question will allow us to answer how observable hadrons are made 
of Nature's elementary and unobservable degrees of freedom.

A closely related problem is dynamical chiral symmetry breaking (DCSB). While hadrons make up most of the visible mass of our universe, this mass cannot be accounted for with the mass of 
light current quarks and perturbatively mass-less gluons. Instead, the interactions between quarks and gluons and between the gluons themselves must be so strong that an effective dynamical 
mass is produced within the hadrons. Only through the generation of such a dynamical mass can the spectrum of experimentally observed hadron masses be explained. 

In applying the theory to understand the hadron's bound-state properties, one immediately realizes that nonperturbative quantum effects play a key role. It is therefore crucial to probe the 
behavior of strong coupling in the infrared domain, so to shed light on confinement confronting theoretical predictions with experimental observables. This is the current objective 
and the motivation for using electromagnetic probes of mesons, protons, and nuclei to measure the rich spectrum of excited and exotic states and the momentum distribution of quarks and 
gluons within them at the upgraded Thomas Jefferson National Accelerator Facility (JLab)~\cite{Holt:2010vj,Aznauryan:2012ba,Accardi:2023chb}, as well as at COMPASS++~\cite{COMPASS:2007rjf}. 
On the other hand, the indirect role that hadron structure plays in most hadron colliders has been  exploited for years, where the aim now is to overcome challenges at the precision level~\cite{LHCHiggsCrossSectionWorkingGroup:2011wcg,PDF4LHCWorkingGroup:2022cjn}. 

In these electron-hadron scattering processes, the scattered particles interact predominantly through the exchange of a virtual photon, which at sufficiently large virtuality serves as a probe to resolve 
quark degrees of freedom inside the hadron. While the hadronic structure resolved at low momentum transfer is different from that in deep-inelastic processes, both can be related via evolution 
equations in QCD. On the other hand, though electron scattering off a fixed hadron target allows us to probe their valence-quark structure, larger collider energies are required to gain access 
to the gluon-field distribution. Since gluons are the force carrier of strong interactions, it is believed that they and their interplay with valence and sea quarks are responsible for the bulk of the 
aforementioned phenomena and puzzles. 

Parton distribution functions (PDFs) are universal, one-dimensional objects that characterize the light-front momentum fraction of partons (quarks and gluons) inside a hadron of a given spin 
configuration.  A PDF can phenomenologically be determined at higher energies through global analyses in QCD for which past (\emph{e.g.} DESY) and present facilities play an important role; see 
Ref.~\cite{Amoroso:2022eow} for a review. Moreover, they provide an indispensable tool for any kind of quantitative phenomenological studies at hadron-hadron colliders, such as the Large Hadron 
Collider (LHC). In particular, the lack of precise knowledge of PDFs is often a limiting factor in reaching high-precision theory predictions at the LHC. The challenge at the future Electron-Ion Collider~(EIC) 
will be to measure appropriate observables that provide information on the spatial distribution and motion of the quarks and gluons in the hadron beyond the 1D structure. Indeed, these longitudinal 
and transverse momentum distributions can be studied with  quantum-correlation functions that can be related to observable scattering amplitudes. Notable amongst them are the transverse 
momentum distributions~(TMDs)~\cite{Boussarie:2023izj} and the generalized parton distributions~(GPDs)~\cite{Kumericki:2016ehc}. With these objects it becomes possible to understand how 
quarks and gluons are distributed in coordinate and momentum space and how spin and angular momentum are carried by them inside mesons and nucleons.  In doing so, several key questions 
in hadron physics and QCD can be addressed: what is the internal multi-dimensional landscape of the nucleon? What is the role of gluons and their  self-interactions in the nucleon? 
What role do collective effects of gluons play in atomic nuclei? How can emergent properties, such as  DCSB  and confinement, be responsible for more than 90\% of the hadron mass?

A slightly different, but related topic, is the possible formation of a very dense and over-occupied system of gluons, carrying a small longitudinal momentum fraction $x$ at large center-of-mass 
energies, which will eventually saturate. Due to a large relative boost factor between the colliding virtual photon and hadron,  pair creation and annihilation of gluons is time dilated in such reactions. 
Since each created gluon is a color source and can therefore create further gluons, one observes a power-like growth of the gluon distribution in this region of phase space. If such a growth is 
extrapolated to infinite high energies, it would lead to the violation of unitary bounds. The growth with energy must therefore slow down at sufficiently high energies. 
According to our current understanding, this takes place through recombination effects in this dense and over-occupied system of gluons, thus slowing down the growth with energy and resulting 
in a phenomenon known as gluon saturation~\cite{Gribov:1983ivg}. The systematic exploration of such a saturated system is expected to be possible in collisions of electrons with large nuclei, 
where densities will be further increased through the nuclear mass number $A$ as $\sim A^{\frac{1}{3}}$, providing the opportunity to investigate the full non-linear dynamics of QCD in a weak 
coupling limit. The color glass condensate~\cite{Iancu:2003xm}  provides an effective description of this saturated regime with many experimental consequences~\cite{Morreale:2021pnn}. 
Furthermore, understanding these systems is also of high importance to control the initial state before the formation of a quark-gluon plasma in relativistic heavy-ion collisions. 

Recent research activities further emphasize the close connection between color confinement and entanglement of microscopic degrees of freedom in the hadron wave function. 
In this context, confinement is understood as the limit of maximal entanglement since colored degrees of freedom cannot exist in isolation, see Ref.~\cite{Beck:2023xhh} for a recent review. 
From this perspective, the DIS process can be interpreted as a sudden quench of the hadronic wave function, due to the interaction with the virtual photon. This leads to a reduced hadronic 
density matrix and corresponding entropy production, which can be measured in experiment \cite{Kharzeev:2017qzs,Tu:2019ouv}.

All of these questions are at the core of the science program of the future Electron-Ion Collider (EIC)~\cite{Accardi:2012qut,AbdulKhalek:2021gbh,AbdulKhalek:2022hcn} at Brookhaven 
National Lab in New York. The U.S. Department of Energy (DOE) granted Critical Decision 3A (CD-3A), \emph{i.e.\/} the project's final design has been approved and its construction has been  
authorized. This collider will be a groundbreaking research machine, pushing the boundaries of our understanding in accelerator science, particle detector design, high-performance computing, 
and beyond.  The most pressing questions that motivated the construction of this new-generation collider are: \medskip

{\setstretch{1.3}
\noindent
$\blacktriangleright$\;  How do quarks and gluons make up nearly all of the visible matter in the universe?

\noindent
$\blacktriangleright$\;   What is the internal three-dimensional landscape of protons and nuclei?

\noindent
$\blacktriangleright$\;   How do the proton's constituent quarks and gluons and their interactions contribute to its spin?
 
\noindent
$\blacktriangleright$\;  Does the interplay of quarks and gluons with the vacuum lead to confinement?
 
\noindent
$\blacktriangleright$\;  How do quark and gluon distributions differ in a proton and in nuclei?
 
\noindent
$\blacktriangleright$\;  Does gluon saturation into a color glass condensate exist?}
\medskip

\noindent
For a more detailed presentation, we refer to the review in Ref.~\cite{AbdulKhalek:2021gbh} and the official EIC page: \url{https://www.bnl.gov/eic/}.


\section{Latin American physicists and the EIC}

Despite the pandemic, the scientific activity dedicated to the planning of the EIC continued and many advances and physics predictions were made~\cite{AbdulKhalek:2021gbh}  to clear the 
path for future ``golden'' physics measurements and to detail the accelerator and detector concepts required to achieve them. Most of the major conferences in hadron physics dealt with recent 
developments around the EIC and a Yellow Report was prepared~\cite{AbdulKhalek:2021gbh}, signed by more than 400 experts in the field from various parts  of the world, including theorists 
and experimentalists from  Argentina (Universidad de Buenos Aires), Brazil (Universidade Federal do Rio Grande do Sul, Universidade Federal de Pelotas, Instituto Tecnol\'ogico de Aeron\'autica),  
Chile (Universidad T\'ecnica Federico Santa Mar\'ia, Universidad Andres Bello) and Mexico (Universidad Nacional Aut\'onoma de M\'exico). 

Faculty and Staff Scientists at Latin American universities and labs who are members of the EIC user group (EICUG)(\url{https://www.eicug.org}) can be found in the EICUG phone book. 
The EICUG is an international affiliation of scientists dedicated to developing and promoting the scientific, technological, and educational goals and motivations for a new high-energy EIC. 
In particular, it acts as a mediator between the different entities and funding bodies of the participating countries thanks to the EIC User Group Steering Committee 
International Representative. This committee contacts the Scientific agencies of each country for various meetings, including meetings with the DOE.
Most Latin American institutions  reported here rely on an institutional representative in the EICUG Council. It is the case for  UAS, UDLAP, UMSNH and UNAM in Mexico, 
as well as ITA, UNICID, UFPel, UNIFESP and UFRGS in Brazil.

More generally, the hadron, nuclear, and particle physicists of this white paper who have participated in research projects related to the EIC are: \vspace*{2mm}
\begin{description}[leftmargin=!,labelwidth=\widthof{\bf Argentina:   } ]
\item [\bf{Mexico\,$:$}]  Adnan Bashir (Universidad de Huelva \& Universidad Michoacana de San Nicol\'as de Hidalgo), Javier Cobos-Mart\'inez (Universidad de Sonora), 
Aurore Courtoy (Universidad Nacional Aut\'onoma de M\'exico), Martin Hentsch-inski (Universidad de las Am\'ericas Puebla), Roger~Hern\'andez-Pinto (Universidad Aut\'onoma de Sinaloa).
\item  [\bf{Brazil\,$:$}]  Arlene Cristina Aguilar (Universidade Estadual de Campinas), Bruno El-Bennich (Universidade Federal de S\~ao Paulo), Tobias Frederico (Instituto Tecnol\'ogico 
de Aeron\'autica), Victor Paulo Gon\c{c}alves (Universidade Federal de Pelotas),  Gast\~ao Krein (Instituto de F\'isica Te\'orica, Universidade Estadual Paulista), Magno V. T. Machado 
(Universidade Federal do Rio Grande do Sul),  Jo\~ao Pacheco B.~C. de Melo (Universidade Cidade de  S\~ao Paulo), Wayne de Paula (Instituto Tecnol\'ogico de Aeron\'autica).
\item  [\bf{Colombia\,$:$}] Fernando Serna (Universidad de Sucre).
\item  [\bf{Argentina\,$:$}]  Daniel de Florian (Universidad Nacional de San Mart\'in), Rodolfo Sassot (Universidad de Buenos Aires).
\item [\bf{Outside Latin America\,$:$}]  Luis Albino (Universidad Pablo de Olavide, Spain), Ignacio Borsa (University of T\"ubingen), Leandro Cieri (Universitat de Val\`encia,  Spain), 
Isela Melany Higuera Angulo (Thomas Jefferson  National Accelerator Facility, USA), Javier Mazzitelli (Paul Scherrer Institute, Switzerland), \'Angel Miramontes (Universitat de Val\`encia, Spain), 
Kh\'epani Raya (Universidad de Huelva, Spain), Farid Salazar (University of Washington, USA), German Sborlini (Universidad de Salamanca, Spain),  Pia Zurita (Universidad Complutense 
de Madrid, Spain).
\end{description}


\section{Scientific activities and current status}

\label{secpastcontribute}

In the following, we summarize the ongoing research activities of the Latin American community in phenomenology,  perturbative and nonperturbative QCD.

\subsection{Contributions to EIC physics within Latin America}
\label{LAEICgroups}

\noindent{\bf In Brazil}, several groups in the State of S\~ao Paulo are very active in hadron phenomenology,  studying spectroscopy, hadrons at finite temperature and in dense matter, 
functional approaches to nonperturbative QCD and light-front quantum field theory.

At the Universidade Estadual de Campinas, the group of Arlene Cristina Aguilar is interested in nonperturbative QCD phenomena and employs Dyson-Schwinger 
equations~\cite{Bashir:2012fs,Roberts:1994dr} to compute the nonperturbative propagators and vertices of QCD~\cite{Aguilar:2022thg,Souza:2019ylx,Aguilar:2018epe,Aguilar:2024ciu,Aguilar:2015bud}. 
It is known that even small quantitative changes in the quark and gluon propagators or in the fundamental quark-gluon, three-gluon, and four-gluon vertices can lead to qualitative changes
in the description of bound states, resonances, and PDFs, which may eventually leave significant experimental signatures.

At the Universidade Federal de S\~ao Paulo, Bruno El-Bennich studies mass generation due to DCSB in QCD. To that end, the general nonperturbative 
structure of the quark-gluon vertex is explored with the Dyson-Schwinger equation and generalized Slavnov-Taylor identities~\cite{Rojas:2013tza,Albino:2018ncl,Albino:2021rvj,Lessa:2022wqc}. 
In the past, the meson and nucleon resonance spectrum, including the Roper and the parity partner of the nucleon, were studied with Poincar\'e covariant Bethe-Salpeter and Faddeev 
equations~\cite{Rojas:2014aka,Segovia:2015hra,Chen:2017pse,Mojica:2017tvh}.  More recently, the group's focus has been on light-front projection of Bethe-Salpeter amplitudes in order 
to compute the parton distribution amplitude (PDA), PDF and TMD of a wide array of mesons including heavy mesons and quarkonia~\cite{Serna:2020txe,Serna:2022yfp,
daSilveira:2022pte,Serna:2024vpn}.

Since pion targets are an experimentally challenging task, whether at JLab or at the EIC, indirect approaches are explored with the Sullivan process by Jo\~ao Pacheco de Melo's group 
at the Universidade Cidade de S\~ao Paulo. In order to probe the pion-structure, off-shell effects in electromagnetic form factors of light pseudoscalar mesons are 
explored~\cite{Choi2019,Leao:2024agy,Burker2023}, which involves the extrapolation of the meson's wave function off their mass-shell. These studies are in view of the pion and kaon 
elastic form factors which will be measured at large momenta and small virtualities at the EIC~\cite{Arlene2019}. 

At the Instituto de F\'isica Te\'orica, the group of Gast\~ao Krein has been investigating the interactions of heavy quarkonia with atomic nuclei.  As nucleons and quarkonia have no valence 
quarks in common, there is no short-range repulsion due the Pauli exclusion principle (responsible for the nucleon-nucleon hard-core repulsion) and the interaction must involve 
gluonic van der Waals forces~\cite{TarrusCastella:2018php}. Such interactions allows us to access a matrix element related to the QCD trace anomaly, a quantum effect that is key to 
our understanding of the origin of the proton's mass and its distribution within the hadron~\cite{Kharzeev:1995ij}. In a  nuclear medium, the strong scalar and vector mean fields enhance 
such  interactions and likely lead to the formation of an exotic nuclear bound state, for which the EIC provides promising perspectives~\cite{Brodsky:1989jd,Krein:2017usp}. 

Tobias Frederico and Wayne de Paula, both at the Instituto Tecnol\'ogico de Aeron\'autica in S\~ao Jos\'e dos Campos, investigate a dynamical continuum formulation 
of generalized TMDs~\cite{dePaula:2022pcb,dePaula:2023ver} in Minkowski space. To that end, they solve the Bethe-Salpeter equation~\cite{dePaula:2020qna} with the concomitant quark 
dressing~\cite{Duarte:2022yur,Castro:2023bij} to take into account the Goldstone boson nature of the pion and kaon. They demonstrated with the  light-front projection of the Bethe-Salpeter 
amplitude that higher Fock components of the hadron states implicitly contribute to their dynamics. For example, at the pion scale, 30\% of its state is distributed beyond the valence component. In the future, they intend  to amend their light-front calculations in Minkowski space with an improved quark-gluon vertex in the corresponding quark Dyson-Schwinger  equation~\cite{Oliveira:2018ukh,Oliveira:2018fkj,Oliveira:2020yac}.

Victor Paulo Gon\c{c}alves of the Universidade Federal de Pelotas has focused during the last two decades on improving the description of inclusive, diffractive and exclusive processes 
in electron-ion collisions~\cite{Goncalves:2000ex,Goncalves:2004bp,Kugeratski:2005gx,Kugeratski:2005ck,Cazaroto:2008qh,Cazaroto:2008iy,Goncalves:2009za,Goncalves:2010vq,
Cazaroto:2010yc,Carvalho:2012xe,Goncalves:2015goa,Goncalves:2015poa,Goncalves:2018pyn}. In recent years, his group has updated its predictions for the diffractive 
processes~\cite{Bendova:2020hkp} and the exclusive production of vector mesons and photons in coherent and incoherent interactions~\cite{Goncalves:2020ywm,Goncalves:2022wzq,
Bendova:2022xhw,Xie:2022sjm,Xie:2024ogs}  considering the kinematical range that will be probed by the EIC and LHeC. Such studies have   demonstrated that the future experimental 
data will be able to improve the description of the QCD dynamics at high energies and constrain the magnitude of saturation effects.   

Magno Machado deals with exclusive particle production and diffractive deep-inelastic scattering at the EIC at the Universidade Federal do Rio Grande do Sul. 
In Ref.~\cite{Fagundes:2022bzw}, the proton structure functions and vector meson production were investigated in $ep$ processes and later extended to nuclear targets in 
$eA$ collisions~\cite{Fagundes:2024jcq}. The very same approach in momentum representation is able to predict the cross section for  exclusive $Z^0$ production~\cite{Peccini:2022quy} 
as well as the time-like Compton scattering~\cite{Peccini:2021rbt}. The diffractive gluon jet production can be described in the context of a QCD dipole picture, considering the $q\bar{q}g$ 
contribution to diffractive deep inelastic scattering. Predictions for such a process at the EIC were presented in Ref.~\cite{Peccini:2020tpj}. Moreover,  Coulomb corrections can be important 
in inclusive and diffractive $eA$ interactions. It was shown in Ref.~\cite{Goncalves:2017kfd} that these corrections to the total cross sections are important at low-$Q^2$ and small values of 
$x$ and are larger for diffractive interactions.  The group also studies entanglement entropy in $ep$ and $eA$ collisions at small-$x$, where this entropy is contrasted to other entropic 
notions, such as the parton entropy calculated in the color glass condensate  formalism and the semi-classical Wehrl entropy~\cite{Ramos:2020kaj}.
\medskip

\noindent{\bf In Colombia}, high-energy and LHC physics dominates the research activities in most physics departments. However, at the Universidad de Sucre, Fernando Serna has focused   
on the one-dimensional structure of hadrons~\cite{Serna:2020txe,Serna:2022yfp,daSilveira:2022pte} and is currently exploring their three-dimensional distributions. From the projections  of 
the Bethe-Salpeter wave functions on the light front, Serna and his collaborators have recently obtained Light-Front Wave Functions (LFWFs) of pseudoscalar mesons. These wave functions 
provide a comprehensive framework to describe probability amplitudes in a more general context. Using these LFWFs, the group has for the first time calculated TMDs and PDFs of the more  
challenging $D$- and $B$-mesons.
\medskip

\noindent{\bf In Mexico}, the hadron physics community is involved in various aspects of hadron structure and spectroscopy. Adnan Bashir's group in Morelia (UMSNH) carries out continuum 
studies through Dyson-Schwinger  equations in QCD to predict observables for Jlab and the EIC. The group has done extensive studies on the quark propagator, the quark-gluon and quark-photon 
vertices as well as the gluon propagator~\cite{Rojas:2013tza,Albino:2018ncl,Albino:2021rvj,Lessa:2022wqc,El-Bennich:2022obe,Ayala:2012pb,Albino:2022efn,Bermudez:2017bpx}. 
An effective field-theory model of QCD based on a vector-vector contact interaction was put forward in Ref.~\cite{Gutierrez-Guerrero:2010waf} and has been studied extensively to compute 
the mass spectrum, decay constants, electromagnetic form factors, transition form factors and charge radii of pseudoscalar and vector meson form factors of light, heavy as well as heavy-light mesons~\cite{Roberts:2010rn,Roberts:2011wy,Bedolla:2015mpa,Bedolla:2016yxq,Gutierrez-Guerrero:2019uwa,Gutierrez-Guerrero:2021rsx,Hernandez-Pinto:2023yin}. The masses of the 
first radial excitations of these mesons and the corresponding diquarks were also computed~\cite{Paredes-Torres:2024mnz}. More realistic studies in collaboration with Kh\'epani Raya yielded 
predictions for the pion and kaon electromagnetic form factors~\cite{Miramontes:2021exi}, transition form factor of pseudo-scalar mesons to two photons~\cite{Raya:2016yuj,Ding:2018xwy} 
pion and kaon form factors in the time-like region~\cite{Miramontes:2022uyi}. 

In Northern Mexico, Javier Cobos-Mart\'inez and his students compute LFWFs of unflavored vector mesons, as well as electromagnetic form factors, PDFs and GPDs
of heavy-light pseudoscalar mesons. They are currently extending these calculations to the distribution functions of diquarks and nucleons with the Dyson-Schwinger equation
approach to QCD.  Also in the North, at the Universidad Aut\'onoma de Sinaloa,  Roger Hern\'andez-Pinto  has recently been interested in phenomenological analyses of 
electromagnetic and two-photon transition form factors of pseudoscalar mesons within an algebraic model. These analyses, in collaboration with Adnan Bashir, Melany Higuera 
and Kh\'epani Raya, predict the pion and kaon form factors for $Q^2$ up to  40~GeV$^2$,  which corresponds to the projected EIC and JLab $Q^2$-range~\cite{Higuera-Angulo:2024oui}.

Martin Hentschinski at the Universidad de las Am\'ericas Puebla contributed recently to the exploration of entanglement entropy in inclusive~\cite{Hentschinski:2021aux} and 
diffractive~\cite{Hentschinski:2023izh,Hentschinski:2022rsa} DIS. Additional research activities comprise phenomenological studies of vector meson production in exclusive photoproduction 
at the LHC~\cite{Bautista:2016xnp,ArroyoGarcia:2019cfl,Hentschinski:2020yfm,Peredo:2023oym}, which are closely connected to corresponding studies at the EIC as well as formal studies 
of high-energy factorization.

At the Universidad Nacional Aut\'onoma de M\'exico, Aurore Courtoy's group  focuses on phenomenological (or global) QCD analyses for the 1-dimensional structure of 
hadrons~\cite{PDF4LHCWorkingGroup:2022cjn, Courtoy:2022kca,Courtoy:2022ocu,Kotz:2023pbu}, with contributions to the CTEQ-Tung et al. (CT) collaboration, e.g. Ref.~\cite{Ablat:2024muy}. 
Global analyses constitute the bridge between data and theory, which is nowadays also supplemented with lattice-QCD studies, in extracting and determining nonperturbative functions 
or characteristics in a data-driven manner. The UNAM group is hence very active in providing predictions for the EIC~\cite{AbdulKhalek:2022hcn} and, together with the Argentinian groups, 
they form the core of global analyses in Latin America. This specific field of expertise is evolving towards precision and accuracy in the determination of the ubiquitous proton PDFs, 
that enter many theoretical predictions for hadron colliders. The group also has  interests in the phenomenological determination of sub-leading twist distributions~\cite{Courtoy:2022kca}, 
as an example of access to spin-dependent 1-dimensional structures without TMD formalism. \\

\noindent{\bf The Argentinian groups} at Universidad Nacional de San Mart\'in and Universidad de Buenos Aires have focused on different aspects of spin physics and hadronization at the EIC. 
These include both the computation of EIC observables beyond next-to-leading order approximation in QCD \cite{Borsa:2022cap,Borsa:2022irn,Borsa:2020ulb,Borsa:2022vvp} as well as 
impact studies based on the foreseen experimental precision \cite{Borsa:2020lsz,Aschenauer:2019kzf}.


\subsection{Latin American physicists outside Latin America}

An important part of the Latin American community is also based outside Latin America, while contributing substantially to building a connection with the region. 
 A Spanish node, composed of  Leandro Cieri (IFIC-Val\`encia), Pia Zurita (Universidad Complutense de Madrid) and German Sborlini (Universidad de Salamanca), is mostly devoted 
to the development of novel computational techniques to obtain high-precision theoretical predictions in high-energy physics~\cite{LTD:2024yrb, Ramirez-Uribe:2024rjg, Aguilera-Verdugo:2020set, 
deLejarza:2024pgk,Ochoa-Oregon:2024zgm}. They are interested in developing novel Quantum Monte Carlo (QMC) methods specifically tailored to address complex problems in this field. 
Moreover, these groups are  currently working on optimizing the calculation of PDFs and fragmentation functions with the aim to reduce computational costs of hadron-collision simulations.

Also in Spain, at the Universidad de Huelva, Kh\'epani Raya has been examining the structure of pseudoscalar mesons with electromagnetic and gravitational form factors, PDFs, and 
GPDs~\cite{Raya:2015gva,Raya:2016yuj,Ding:2018xwy,Ding:2019qlr,Raya:2021zrz,Xu:2023izo}. His group has recently proposed new methods for deriving PDFs and GPDs of pions and
kaons employing probabilistic methods~\cite{Xu:2023bwv,Lu:2023yna}. The Dyson-Schwinger and Bethe-Salpeter equation framework provides access to a variety of pseudoscalar meson  
distributions, and a similar degree of sophistication is expected for the nucleon in upcoming years. These techniques have also allowed for the computation of accurate hadronic light-by-light 
contributions to the anomalous magnetic moment of the meson, which was done in collaboration with \'Angel Miramontes at the Universitat de Val\`encia and Adnan Bashir~\cite{Miramontes:2021exi,
Raya:2019dnh}. They also resulted in realistic predictions for space-like~\cite{Miramontes:2021exi} and time-like~\cite{Miramontes:2022uyi} electromagnetic form factors of the pion and kaon. 
Melany Higuera, now at Jefferson Lab, and Luis Albino of the Universidad Pablo de Olavide in Sevilla have developed algebraic models for LFWFs and derived therefrom pion and kaon PDFs 
and GPDs~\cite{Albino:2022gzs}. Luis Albino also contributed to the derivation of the nonperturbative quark-gluon vertex structure~\cite{Bermudez:2017bpx,Albino:2018ncl,Albino:2021rvj}.

Farid Salazar is working at the Institute for Nuclear Theory, University of Washington, on various aspects of gluon saturation within color glass condensate effective theory. His recent efforts are focused on 
addressing the consistent resummation of high-energy and Sudakov/threshold-type logarithms in deep inelastic scattering in the saturation regime \cite{Caucal:2023fsf,Caucal:2024cdq,Caucal:2024nsb}, 
as well as the phenomenology of vector meson production in high-energy photon-nucleus reactions in ultra-peripheral heavy-ion collisions \cite{Mantysaari:2022sux,Mantysaari:2023prg,Mantysaari:2023xcu}. 
He is also interested in elucidating the correspondence between different formalisms for multiple scattering in nuclei \cite{Fu:2023jqv,Fu:2024sba}.

Javier Mazzitelli, at the Paul Scherrer Institute in Switzerland, works on the computation of higher-order QCD corrections for collider phenomenology. His recent works mainly deal with Higgs boson and 
heavy-quark phenomenology~\cite{Grazzini:2018bsd,Buonocore:2023ljm}. Besides computing corrections at fixed-order in perturbation theory, he also developed methods to match higher-order calculations 
to parton showers~\cite{Mazzitelli:2020jio,Mazzitelli:2024ura} in order to provide accurate multi-purpose Monte Carlo event generators, a cornerstone of experimental analyses at colliders.

Ignacio Borsa has mainly focused on spin physics at the EIC, including the computation of higher-order QCD corrections for polarized processes~\cite{Borsa:2020ulb,Borsa:2020yxh,Borsa:2022cap,
Borsa:2022irn} and, more recently, on the matching of higher-order corrections to parton showers~\cite{Borsa:2024rmh}. Part of his work has also been centered in the determination of helicity 
parton distributions functions beyond next-to-leading order~\cite{Borsa:2024mss}.


\section{The building of a community}
\label{sec:community}

In 2019, Martin Hentschinski co-organized the international ``Workshop on Forward Physics and  QCD at the LHC, 
the future Electron-Ion Collider, and Cosmic Ray Physics" in the city of Guanajuato. This small meeting aimed at connecting Mexican graduate students with international 
researchers in this area. The ``19th International Conference on Hadron Spectroscopy and Structure'' (HADRON 2021) was held online at the Universidad Nacional Aut\'onoma de M\'exico, 
co-organized by the local EICUG member Aurore Courtoy. 

Bruno El-Bennich chaired the tenth international conference on the ``Physics Opportunities at an Electron-Ion Collider'' (POETIC 2023) at the Instituto Principia in S\~ao Paulo, supported by 
the ICTP South American Institute  for Fundamental Research (ICTP-SAIFR), Jlab and Brookhaven National Lab, while Tobias Frederico was the chair of the 2023 edition of the Light-Cone 
Conference Series ``Light-Cone 2023: Hadrons and Symmetries'', hosted by the Centro Brasileiro de Pesquisas F\'isicas (CBPF) in Rio de Janeiro. Most recently,  Adnan Bashir and 
Bruno El-Bennich co-organized the workshop  ``From Quarks and Gluons to the Internal Dynamics of Hadrons'', while Tobias Frederico  was the co-organizer of ``Elucidating the 
Structure of Nambu-Goldstone Bosons", both at the Center for Frontiers in Nuclear Science, Stony Brook University.  The primary objective of these in-person workshop was to discuss 
open questions related to momentum distributions, form factors, masses, and other observables that are of paramount interest to  the EIC community.

In July 2022 and June 2024, Martin Hentschinski co-organized two meetings, ``Saturation and Diffraction at the LHC and the EIC" and ``Diffraction and gluon saturation 
at the LHC and the EIC", at the European Center for Theoretical Studies in Nuclear Physics and Related Areas ECT* in Trento, Italy. The EIC was also a relevant topic at the ``International workshop 
on the physics of Ultra Peripheral Collisions (UPC 2023)", organized in December 2023 in Playa del Carmen, Mexico. 

The QCD4EIC workshop series is jointly organized by the Buenos Aires group (Daniel de Florian  and Rodolfo Sassot) and Werner Vogelsang of T\"ubingen University at the  Center for Frontiers
in Nuclear Science at Stony Brook.  Those workshops were attended by members of the Latin  American community ({\it e.g.} Aurore Courtoy, Ignacio Borsa). These workshop
series revolved around precision fixed-order calculations of polarized and unpolarized hard-scattering cross sections at the EIC, QCD threshold resummation for future EIC cross 
sections and their impact on phenomenology, extraction techniques for PDF and FF extractions  from future EIC  data in the context of global analyses, the prospect for ``joint" global 
analyses that aim at simultaneous extractions of PDFs and FFs, and high-$p_T$ observables at the EIC and their interplay with TMD physics.

Other activities include a lecture series Adnan Bashir gave at the ``38th Annual Hampton University Graduate Studies'' (HUGS) Summer School  at Jlab and at the  
``Summer School on the Physics of the EIC'' at the Center for Frontiers in Nuclear Science, while Fernando Serna spent a month as a visitor of the EIC Theory Institute
at Brookhaven National Lab. Other Latin American physicists are strongly encouraged to take advantage of the EIC Theory Institute's visitor program in the 
future.\footnote{\,Application instructions can be found at: \url{https://www.bnl.gov/eic-theory/}}


\section{Current projects and future Latin American contributions to the EIC}

In Section~\ref{secpastcontribute} we made the case that the Latin American EIC community is steadily increasing and has contributed significantly to the science of the EIC. 
All groups mentioned earlier will continue to contribute to the EIC physics program on various fronts, be it in phenomenology, in field theoretical developments and
computational advances. We expect increasing activity in the fields of prime interest to the international EIC community, namely the \emph{origin of visible matter\/}, 
\emph{quark and gluon confinement\/}, \emph{precision 1D and 3D imaging of pions, protons and nuclei\/}, \emph{the proton spin puzzle\/}, \emph{gluon saturation\/} and 
\emph{ quark and gluon  distributions in nuclei\/}. 

The contributions of the Latin American theoretical community span from nonperturbative QCD approaches and perturbative QCD tools to global QCD analyses. 
 As such, the present document represents a Latin American counterpart of the initiative for an EIC Theory Alliance~\cite{Abir:2023fpo}. The latter advocates for a
coordinated effort involving theory groups for which differing expertise is needed. The future Latin American contributions can be grouped thematically as follows.


\subsection{Three-dimensional tomography of hadrons}

The long-term objective is to improve upon earlier studies of the three-dimensional distributions of light and heavy mesons and to extend them to make predictions for the nucleon's 
TMDs, GPDs and gravitational form factors beyond parametrizations and algebraic models. Within the Latin American hadron physics community, functional methods have proven to 
be a popular approach to compute Euclidean Green functions and to study the mass spectrum and structure of mesons and baryons, their electromagnetic and electroweak form 
factors and distribution functions. Much effort revolved around light  and heavy mesons in the pseudoscalar and vector channel and will be extended to baryons with modern versions
of Faddeev wave functions.  \\

\noindent{\bf Parton Distribution Functions and Electromagnetic Form Factors.} 
While the long-term goal is to make competitive predictions for the nucleon GPD and related gravitational form factors, the precision of truncation schemes~\cite{Albino:2021rvj,Albino:2018ncl,
Gao:2024gdj} is important as the Bethe-Salpeter wave functions of mesons are significantly more sensitive to them than their mass poles. 
The corresponding LFWFs are the starting point to derive any kind of distribution functions~\cite{Serna:2024vpn}  and systematic improvements will be initially be applied to the PDF of unequal 
mass states, such as the $D$ and $B$ mesons or the topical $D_{s0}$ and $D_{s1}$ mesons, where the leading ladder truncation fails. In the long term, considerable activity will also be dedicated 
to the calculation of distribution functions in Minkowski space~\cite{dePaula:2022pcb,dePaula:2023ver}, employing Nakanishi representations of Bethe-Salpeter amplitudes and spectral 
representations for quark and gluon propagators via inversion procedures.  

Including higher-truncation contributions will also improve the precision of the pion and kaon time and space-like electromagnetic form factors~\cite{Miramontes:2022uyi,Choi2019} and 
lead to a better understanding of the mass-generation mechanism. The pion, with its nearly massless nature, offers a clear demonstration of how chiral symmetry operates, revealing 
measurable effects in its electromagnetic form factor. For the kaon, the strange-quark content introduces a more significant role for the Higgs mechanism, providing a complementary 
perspective. Phenomenological ans\"atze to form factor extraction will be provided and can be tested once data are released. The analysis of, e.g., axial-vector form factors using 
the combined Dyson-Schwinger and Bethe-Salpeter equations approach can be relevant for observables such as the anomalous magnetic moment of the muon. The Latin American 
groups also  play an important role in the phenomenological determination of the 1D structure of mesons and nucleons  through global analyses of PDFs, and will continue to do so --- see 
Section~\ref{subsec:comple_HEP} below.
\\

\noindent{\bf Transverse Momentum Distributions.} Several Latin American groups have derived TMDs from the LFWF of the leading meson's Fock state and the above observation 
about truncation schemes holds for them too. That is, higher-order effects in the interaction kernels have important consequences for DCSB and both, the PDF and TMD, at least of light 
mesons, become generally broader as a function of $x$. The analytic behavior of the transverse-momentum dependence of the TMDs, in particular the inflection point at a typical 
hadronic scale in their steady decrease, must be understood in terms of the competition between the Higgs mechanism and DCSB before one engages on the much more challenging 
task of computing the nucleon TMD. An important issue to be addressed are higher Fock-state contributions, as the light-front projection of the pion's Bethe-Salpeter amplitude shows 
that only 70\% of the pion stems from its valence components at a low hadronic scale~\cite{Duarte:2022yur,Castro:2023bij}. Likewise, baryon-number conservation of the LFWF is 
not conserved with the leading Fock states of the pion~\cite{Serna:2024vpn}. These efforts will complement lattice QCD studies on quasi-PDFs and upcoming experimental data.
\\

\noindent{\bf Generalized Parton Distributions and Gravitational Form Factors.}   Direct measurements of gravitational form factors (GFFs)  are nearly impossible due to the weakness 
of the gravitational interactions. However, processes like deeply virtual Compton scattering (DVCS) provide an indirect approach to accessing GFFs. In the Latin American community, 
GPDs and gravitational form factos (GFF) of pseudoscalar mesons have so far been explored with algebraic models~\cite{Raya:2021zrz,Albino:2022gzs} that mimic the nonperturbative  
infrared and perturbative ultraviolet behavior of quark propagators and Bethe-Salpeter amplitudes while preserving chiral symmetry. The calculation of  valence quark GPDs of a variety 
of mesons within the Dyson-Schwinger equation approach employs the overlap representation of the LFWF in the Dokshitzer-Gribov-Lipatov-Altarelli-Parisi (DGLAP) evolution region, 
i.e. where $x$ is larger than the skewness parameter $\xi$. First results are expected soon. Clearly, the efforts to better control truncation effects and understand 
the impact of DCSB on one- and three-dimensional momentum distributions serves as benchmarks for the nucleon GPD predictions for the EIC. 

The Laboratory Directed Research and Development team at Jefferson Lab, of which Melany Higuera is a member, has focused on exploring the gravitational structure of  the proton 
using innovative empirical methods. The LDRD efforts combine the expertise of experimental physics, lattice QCD theory, and computational science to enhance data analysis with robust 
theoretical constraints and advanced fitting methods, including neural networks. These efforts aim to reduce model dependence and improve the accuracy of results, offering deeper insights 
into the proton's mechanical properties.
\\

\noindent{\bf Fragmentation Functions.}
 Elementary $q\to \pi, K, \rho \ldots$, fragmentation functions can been obtained from a cut-diagram based on the optical theorem and are  related to the meson's PDF via a Drell-Levy-Yan 
 relation exploiting crossing and charge symmetry. As one must consider the possibility that the fragmenting quark produces a cascade of mesons, this elementary probability distribution 
 defines the ladder-kernel of a quark-jet fragmentation equation, similar to the quark jet-model of Fields and Feynman~\cite{Field:1976ve}, 
 from which one obtains the full  fragmentation function for a given meson~\cite{daSilveira:2024ztt}. Future studies will address the more challenging gluon fragmentation functions, 
 as they imply higher-order diagrams and the precise knowledge of the nonperturbative quark-gluon vertex~\cite{Aguilar:2018epe,Aguilar:2024ciu,Oliveira:2018ukh,
 Oliveira:2018fkj,Oliveira:2020yac,Rojas:2013tza,Albino:2018ncl,Albino:2021rvj,Bermudez:2017bpx}  in the hadronization of energetic gluons. Latin American groups have also 
 contributed  the phenomenological determination of meson fragmentation functions through global analyses, and will continue to do so, see Section~\ref{subsec:comple_HEP} below.


\subsection{Charmonium photo-production}

The precise determination of the $J/\psi$ and $\Upsilon$ photo-production at low energies~\cite{Duran:2022xag} is motivated by theoretical studies that indicate that the  near-threshold 
production of heavy quarkonium is sensitive  to the trace-anomaly contribution to the nucleon mass. Moreover, photon induced interactions can also be used to study the production and properties  
of  exotic charmonium-like states~\cite{vicwer,vicmar,nosexotico,Goncalves:2018hiw}, which are a class of hadrons that decay to final states that contain a heavy quarks and antiquarks 
but cannot be easily accommodated in the remaining unfilled states in the $c\bar{c}$ level scheme. It is important to improve the description of the photo-production of exotic states and to 
derive predictions for its production, considering $ep$ and $eA$ collisions at the EIC motivated by narrow, hidden heavy-quark pentaquark states $P_c(4312)$, $P_c(4440)$ and $P_c(4457)$. 
The latter were reported by the LHCb Collaboration~\cite{lhcb_penta}. Particle production via $\gamma \gamma$ interactions in future electron-ion collisions was recently discussed 
in detail in Refs.~\cite{Babiarz:2023cac,Francener:2024eep,Bertulani:2024vpt} and indicates that the properties of QCD and QED bound states could be constrained at the EIC and LHeC.

While model-dependent, the $J/\psi$ cross section can also serve to extract the gluon GPD of the nucleon. This can be achieved at threshold where the kinematics has a large 
skewness parameter $\xi$, leading to the dominance of the spin-2 contribution over higher-spin twist-2 operators~\cite{Guo:2021ibg}. In this limit one can express the gluon GPD 
in terms of the corresponding energy momentum tensor of the proton, though more sophisticated, nonperturbative calculations can verify the assumptions used in the factorization. 
In particular, a reliable extraction of the gluon GPD is sensitive to the $J/\psi$ photo-production amplitude and should not rely on vector-dominance models or simplistic production 
mechanisms~\cite{Xu:2021mju}. An important input is provided by realistic  $J/\psi$ and $\eta_c$ distribution amplitudes~\cite{Serna:2022yfp}.


\subsection{Complementarity with other high-energy colliders}
\label{subsec:comple_HEP}

\noindent{\bf Small-$x$ physics.} 
An important future research activity should combine data collected both at the EIC and at the LHC to identify and pin down the driving factors of QCD high energy evolution, so that one 
can address the relevance of different types of high density corrections. To  answer this question, a combination of both EIC and LHC data is thus necessary. While data collected at the EIC 
will allow to determine the hadronic structure with high precision, the LHC data provides on the other hand the necessary reach in energy to test QCD evolution towards higher energy or 
lower fractions of $x$.  A particular interesting class of processes are ultra-peripheral collisions at the LHC, since they allow to study photon-proton and photon-nucleus reactions, 
similarly to the collisions that will be studied at the EIC.\\

\noindent{\bf Data-driven analyses of collinear structures.} 
While the model approaches described above aim to {\it predict} distribution functions from a low scale, global analyses {\it extract} PDFs from data through solving inverse problems based 
on a perturbative QCD description of the observables. Collinear PDFs are determined from diverse datasets, spanning a range of processes and energy scales. The EIC, which will probe 
the large-$x$ and intermediate-$Q^2$ regions, is expected to improve PDF determinations in what is currently referred to as ``extrapolation regions." PDFs at large $x$, on the other hand, 
are heavily influenced by nonperturbative QCD effects due to the kinematical vicinity with the elastic limit and the transition between QCD's two regimes --- features that serve as crucial 
initial conditions for PDF extraction. While the precision and accuracy of collinear PDFs are now unprecedented, global analyses face challenges typical of big-data studies, particularly 
in the robust quantification of uncertainties.
The latter is an active research topic, with significant current and upcoming contributions from the Latin American community. New methodologies are being developed to better account 
for uncertainty sources, with applications to the LHC\footnote{See the recent discussions at the PDF4LHC 2024 meeting, \url{https://indico.cern.ch/event/1435677/}.} and EIC data, 
but also for the intersection with lattice data. Indeed, lattice QCD is now able to provide information on quantities that are close to distribution functions. Future advancements in global 
analyses will hence triangulate between experimental data, lattice data and theory precision.

 
\subsection{Precision calculation}

With a solid experience in multi-loop calculations, high-precision predictions for EIC observables are of great interest to the Latin America EIC community. The initial aim is to reach NLO 
followed by NNLO accuracy in the calculation of relevant partonic processes. Parton-level results will be combined with ion/heavy ions models in order to compute the full contributions to cross-sections 
and other observables to be measured at the EIC. Several groups in Spain and Argentina are experts in phenomenological analysis and plan to explore novel observables, such as  $X$ 
(hadron for instance) + direct photon production, while closely  interacting with the experimentalists to implement relevant measurements. 


\subsection{Quantum information}
 
A very recent but rapidly growing research area addresses the interface of nuclear physics and quantum information to which several groups will contribute in the future. 
A particularly interesting topic is entanglement in high energy reactions, as it tests fundamental aspects of quantum mechanics in an highly relativistic environment but also promises 
to be instrumental in probing non-trivial aspects of strong interactions in the proton wave function, for instance.


\subsection{Outreach, innovation, diversity and benefits for society} 

The Latin American community brings invaluable diversity to the EIC user group while maintaining a robust scientific foundation. This international component is complemented by strong diversity, 
equity, and inclusion (DEI) policies implemented at both, national and local levels, in each represented country, in some cases via affirmative action in faculty hiring and by fellowship-granting 
agencies. Engagement and collaborations (\textit{vinculaci\'on} in Spanish, \textit{integra\c{c}\~ao} in Portuguese) play a pivotal role in fostering multidisciplinary interactions, whether with experimentalists, 
computer scientists, or statisticians, among others. These interactions have positively influenced funding agencies by demonstrating the broader societal impact of our work.

Disseminating (\textit{divulgaci\'on cient\'ifica} in Spanish, \textit{divulga\c{c}\~ao cient\'ifica} in Portuguese) EIC-related physics has significantly increased the visibility of ongoing activities (see 
Section~\ref{sec:community}) while also strengthened the community itself. This has been particularly evident through the active involvement of students and early-career researchers, 
who are integral to the future of the field. Outreach programs ({\it difusi\'on} in Spanish, \textit{expans\~ao} in Portuguese) at universities provide activities and expertise to people outside 
the university community, often in collaboration with local municipalities and schools, while teachers are encouraged to attend particle physics summer schools at the ICTP-SAIFR, for instance. 
In that context, it is worthwhile to mention that CERN schools are offered to Latin American high-school teachers in Spanish and Portuguese and can serve as an inspiration for similar outreach 
programs at the EIC.


\section{Perspectives and challenges}

The Latin American groups have mostly worked on their own or within limited national collaborations (Morelia--Sinaloa, Morelia--UNAM, S\~ao Paulo-Pelotas, for example) or international 
ones (Morelia--S\~ao Paulo, Sucre--S\~ao Paulo), while individual collaborations with colleagues in Europe and in the U.S. are more common. Thus far, lack of logistics and coordination 
have not permitted the creation of schools and workshop series specifically dedicated to the EIC. Joining efforts will be valuable given the large number of talented young students and 
early-career researchers involved in the research activities of most groups mentioned in Section~\ref{secpastcontribute}. 

To that end, we shall resume the alternating workshop series, ``Many manifestations of nonperturbative QCD'' and ``Nonperturbative Aspects of Field Theories'', which used to take
place in S\~ao Paulo and Morelia and were interrupted by the pandemic. This workshop series ought to be extended to regions where EIC physics in particular, and hadron physics in 
general, is still in its infancy, for example in Colombia. We also foresee a continuity in contributing to international meetings at the Center for Frontiers in Nuclear Science in Stony Brook, 
as well as at the Institute for Nuclear Theory in Seattle, the European Center for Theoretical Studies in Nuclear Physics and  Related Areas in Trento, at the ICTP-SAIFR in S\~ao Paulo 
and at the Mesoamerican Centre for Theoretical Physics in Chiapas. 

The financial support of the Inter-American Network of Networks of QCD challenges (I.ANN-QCD) has been very valuable in the past, though this U.S. initiative ought to be supplemented 
by Latin American agencies. A similar role is played by the visiting program of the EIC Theory Institute at Brookhaven National Lab. Scientific visits, student and postdoc exchanges 
should therefore be coordinated within the framework of bilateral or multilateral agreements. 
In this context, the SPRINT calls of the  Funda\c{c}\~ao de Amparo \`a Pesquisa do Estado de S\~ao Paulo (FAPESP) can serve as a blueprint for future initiatives~\cite{SPRINT}: 
proposals for inter-institutional or thematic collaborations ought to be submitted to the national funding agencies, for instance FAPESP and Conselho Nacional de Desenvolvimento 
Cient\'ifico e Tecnol\'ogico (CNPq) in Brazil, the SECIHTI (Secretar\'ia de Ciencia, Humanidades, Tecnolog\'ia e Innovaci\'on) in M\'exico and the Consejo Nacional de Investigaciones 
Cient\'ificas y T\'ecnicas (CONICET) in Argentina. Such international calls for projects lasting two to three years would be extremely helpful to maintain long-term collaborations in Latin America. 
Visits can in principle also be financed by the Centro Latino-Americano de F\'isica (CLAF), though its financial situation appears to be precarious. 

In the coming years, we foresee an increase of publications in peer-reviewed journals related to and motivated by the EIC. Those will be supplemented by  talks in dedicated international
workshops and conferences, as well as lectures in schools, that will take place in Latin America in the coming years. To that end, we estimate the financial needs as follows. \\
\begin{itemize}
    \item Based on the experience of organizing POETIC~2023 in S\~ao Paulo, the Latin American EIC community anticipates that each large event will cost about US\$30,000, comprising 
    of about 40-50 invited speakers and including accommodation, airfares, venue and coffee breaks. Medium-size events attended by fewer speakers, for 
    example summer schools for students, are expected to cost about US\$15,000 per week (we divide the cost of typical US-based schools by two) for 50-60 students, in addition to a similar
     amount to cover the lodging and travel costs of lecturers.
     Hence,  between schools and conferences, an annual budget of about \$US50,000-60,000 is necessary for the organization of Latin American EIC events that would consolidate the community.\\
    \item The I.ANN-QCD supports projects between Latin American and U.S. institutions, as well as the participation of Latin-America based speakers in conferences and 
    workshops  in the U.S. Supplementing this I.ANN-QCD  support  would amount to about US\$30,000 per year. \\
    \item The Center for Frontiers in Nuclear Science, dedicated to the EIC experimental program, provides 25\% of funding for joint postdoctoral fellowship. However, it has proven difficult 
    in the past to justify and/or find complementary support from Latin American institutions. Any effort in that direction will be useful.\\
    \item Student exchanges and scientific visits of collaborators have mostly been financed with individual grants. We expect and will support more coordinated actions involving State and 
    Federal agencies  in Latin America that result in calls for bi- or multilateral projects dedicated to EIC physics. 
\end{itemize}
\medskip

\noindent 
In summary, the Latin American hadron physics community is clearly  excited about the EIC science program that aims to answer fundamental questions about the structure and properties 
of matter. The involvement goes beyond individual engagement and regional or national collaborations have evolved into international ones. It is more than likely that frequent EIC meetings 
in Latin America and in the United States, attended and often organized by Latin Americans, contributed to this evolution. The seeds have been sown for fruitful, innovative and lasting 
collaborations and exchanges between South and North America, and several Latin American physicists in Europe also focus on EIC physics. We are hopeful that the current level of 
mobility will increase and also allow for more student exchanges, while we also aim at increasing Latin American membership in the EICUG and participation in EIC science politics.

The past decades have witnessed  the emergence of very active experimental groups, though often involved in high-energy experiments at the LHC, SuperKEKB, Pierre Auger Observatory, 
Fermilab, JUNO or ANDES, amongst others. Recent initiatives, in particular in Mexico, involve experimental colleagues in discussing the possibilities to contribute to software development, 
electronics, and detector design and to propose experiments once the EIC is operative. It is therefore up to the Latin American EIC community to win over keen experimentalists seeking 
challenges and exciting new paths for the decades to come.


\section{Acknowledgements}

We acknowledge financial support by the following Science and Research Agencies: \smallskip \\
 Funda\c{c}\~ao de Amparo \`a Pesquisa do Estado de S\~ao Paulo (FAPESP),  grant nos.~2023/00195-8 and 2018/25225-9,
Funda\c{c}\~ao de Amparo \`a Pesquisa do Estado do Rio Grande do Sul (FAPERGS),  
Conselho Nacional de Desenvolvimento Cient\'ifico e Tecnol\'ogico (CNPq), grant nos.~309262/2019-4,  313030/2021-9, 409032/2023-9,  401565/2023-8  and 310763/2023-1,
Coordena\c{c}\~ao de Aperfei\c{c}oamento de Pessoal de N\'ivel Superior, grant no.~88881.309870/ 2018-01,
Instituto Nacional de Ci\^encia e Tecnologia: F\'isica Nuclear e Aplica\c{c}\~oes (INCT-FNA), grant no.~464898/2014-5,
Consejo Nacional de Investigaciones Cient\'ificas y T\'ecnicas (CONICET),
Consejo Nacional de Humanidades, Ciencias y Tecnolog\'ias (CONAHCyT), project  CBF2023-2024-3544, CBF2023-2024-268 and Ciencia de Frontera 2019 No.~51244 (FORDECYT-PRON-ACES),
Universidad Nacional Aut\'onoma de M\'exico, grant no.~DGAPA-PAPIIT IN111222,
Coordinaci\'on de la Investigaci\'on Cient\'ifica of the Universidad Michoacana de San Nicol\'as de Hidalgo, grant no.~4.10,  
Generalitat Valenciana GenT Excellence Programme, grant no. CIDEGENT/2020/011, ILINK22045, Programme PROMETEO of the Generalitat Valenciana, grant no.~CIPROM/2022/66.,
``Atracci\'on de Talento'' of the Comunidad de Madrid, grant no.~2022-T1/TIC-24024,
Agencia Estatal de Investigaci\'on, grant nos.~PID2020-113334GB-I00, PID2023-151418NB-I00 and PID2022-140440NB-C22,
Regional Andalusian project P18-FR-5057, Deutsche Forschungsgesellschaft (DFG) research unit FOR 2926.


\bibliography{biblio}

\end{document}